\documentclass[aps,twocolumn,showpacs,preprintnumbers,amsmath,amssymb]{revtex4}
\usepackage{graphicx}
\usepackage{bm}
\usepackage{dcolumn}

\def\nat#1#2#3{Nature {\bf #1}, #2 (#3)}

\def\sc#1#2#3{Science {\bf #1}, #2 (#3)}
\def\prv#1#2#3{Phys. Rev. {\bf #1}, #2 (#3)}
\def\rmp#1#2#3{Rev. Mod. Phys. {\bf #1}, #2 (#3)}
\def\prl#1#2#3{Phys. Rev. Lett. {\bf #1}, #2 (#3)}
\def\pra#1#2#3{Phys. Rev. A {\bf #1}, #2 (#3)}

\def\epjd#1#2#3{Eur. Phys. J. D {\bf #1}, #2 (#3)}
\def\njp#1#2#3{New J. Phys. {\bf #1}, #2 (#3)}

\def\jpamg#1#2#3{J. Phys. A: Math. Gen. {\bf #1}, #2 (#3)}
\def\jpb#1#2#3{J. Phys. B: At. Mol. Opt. Phys. {\bf #1}, #2 (#3)}

\def\pla#1#2#3{Phys. Lett. A {\bf #1}, #2 (#3)}

\def\jltp#1#2#3{J. Low Tem. Phys. {\bf #1}, #2 (#3)}

\def\noi{\noindent}
\def\bc{\begin{center}}
\def\ec{\end{center}}
\newcommand{\bea}{\begin{equation}}
\newcommand{\eea}{\end{equation}\noi}
\newcommand{\ber}{\begin{eqnarray}}
\newcommand{\eer}{\end{eqnarray}\noi}
\begin{document}
\title{Excess energy of an ultracold Fermi gas in a trapped geometry}
\author{Shyamal Biswas$^1$}\email{sbiswas.phys.cu@gmail.com}
\author{Debnarayan Jana$^1$}
\author{Raj Kumar Manna$^{2}$}
\affiliation{$^1$Department of Physics, University of Calcutta, 92 APC Road, Kolkata-700009, India \\
$^2$Department of Physics, Indian Institute of Technology-Madras, Chennai-600036, India}

\date{\today}

\begin{abstract}
We have analytically explored finite size and interparticle interaction corrections to the average energy of a harmonically trapped Fermi gas below and above the Fermi temperature, and have obtained a better fitting for the excess energy reported by DeMarco and Jin [Science $\textbf{285}$, 1703 (1999)]. We have presented a perturbative calculation within a mean field approximation.
\end{abstract}

\maketitle
\section{Introduction}
Observation of quantum degeneracy of $^{40}$K atoms \cite{jin} in harmonically trapped geometry made the study of ultracold fermions an hunting ground to the experimentalists \cite{jin1,fermi-p,f-superfluidity,cv,thomas,nascimbne} and theoreticians \cite{pitaevskii-rmp2} within the last ten years. In the remarkable experiment, DeMarco and Jin measured the energy of harmonically trapped weakly interacting $^{40}$K atoms for different temperatures ($T$) \cite{jin}. They plotted the deviation of the average energy of a single $^{40}$K atom from its classical part ($3kT$). This deviation (excess energy) particularly below the Fermi temperature ($T_F\sim0.6~\mu\text{K}$) clearly indicates that $^{40}$K atoms obey Fermi-Dirac statistics. They also compared their experimental data with the ideal gas prediction. Although this prediction matches well with the experimental data yet the measured energy is a little higher, and it maintains a systematic narrow gap with it in particular for $T/T_F\lesssim1$.

It is obvious that confinement in a smaller region and repulsive interparticle interaction increase the average energy of a particle. Since the thermodynamic limit is not well satisfied for our system, as its size ($\sim\text{mm}$) and number ($\sim10^6$) of particles are small \cite{jin,jin1}, finite size effect may contribute in removing the narrow gap between the theory and experiment. DeMarco and Jin produced a nearly ideal Fermi gas \cite{jin}. But, interparticle interaction can not absolutely be neglected in their experiment, and it may also contribute in removing the gap \cite{jin1}.

Besides the study of strongly correlated electron (Fermi) gas, weak interaction effects on Fermi gas have also been studied for a long time \cite{lee}. Finite size effect on an ideal Fermi gas was also studied a few decades ago \cite{barma}. While the Fermi gas for those cases are essentially homogeneous, the Fermi gas of our interest is inhomogeneous being trapped by an inhomogeneous magnetic field. Finite size effect was explored for such a system by the authors of Ref.\cite{finitesize-fermi} in a way Grossmann and Holthaus did for Bose gas \cite{finitesize}. They considered the finite size effect on our system by adding the zero point energy of the oscillations and a surface contribution (resulted from the error of considering discrete energy levels continuum) to the bulk density of states \cite{finitesize-fermi}. They eventually showed that these two corrections cancel each other in the energy expression to the first order in angular frequency. Consequently, they did not get considerable finite size correction at all. However, apart from these, finite size effect may also be there in measuring temperature \cite{temperature}, and in calculating the chemical potential as well. Incorporating these features one can get considerable finite size correction to the first order in angular frequency. Along with the finite size effect we also consider the contribution of weak interparticle interaction in energy for entire range of temperature. Although the interest of exploring the thermodynamic behavior of an ultracold Fermi gas has been moved to strongly correlated regime \cite{ho,perali,carr,thermodynamics,bulgac,hu2} yet the same for weakly interacting regime has not surprisingly been studied except that for $T\rightarrow0$ \cite{T-F,M-F}, and for $T\ne0$ with a particular interest in phase separation of its multi-components \cite{salasnich}.

In this paper we will incorporate the temperature dependence of finite size and weak interaction effects on a harmonically trapped ultracold Fermi gas in a perturbative manner. Our calculation will begin in section-II with a mean field Hamiltonian where we will consider short ranged interaction among the particles \cite{M-F,T-F}. We will write the average occupation number for the interacting fermions in a self consisted way Giorgini, Pitaevskii and Stringari wrote for an interacting Bose gas \cite{giorgini}. We will obtain a considerable shift in chemical potential due to the interparticle interaction. In section-III we will incorporate finite size effect by considering shifts in temperature and chemical potential apart from the previous treatment \cite{finitesize-fermi}. In section-IV we will evaluate the interaction energy. We will get the excess energy in a closed form of fugacity (or chemical potential) in section-V. Then we will obtain an approximate chemical potential in a closed form of temperature. This approximate chemical potential will make the excess energy in a closed form of temperature. Then we will replot the experimental data for evaporation ramp indicating the temperature dependence of the number of particles \cite{jin}, and will fit the data points. We will eventually plot this excess energy for spin $\uparrow$ and $\downarrow$ particles by considering the fitting formula for the evaporation ramp, and will compare our analytic result with the experimental data \cite{jin}. In section-VI we will compare the interaction energy with the oscillator energy. As a corollary, we will obtain a considerable shift in Fermi energy due to the interparticle interaction. Finally, we will compare the interaction and finite size corrections.
\section{Mean field theory}
Let us consider a weakly interacting trapped Fermi gas of $N$ number of 3-D anisotropic harmonic oscillators each having spin $1/2$. Each single particle state $\{\textbf{p},\textbf{r}\}$ is described by a specific momentum ($\textbf{p}$) and a specific position ($\textbf{r}$) from the center of the trap, and each state of course has $2$ spin degeneracy ($\sigma=\uparrow,\downarrow$). We consider unequal number of particles in the two spin states. It is convenient to begin with an isotropic trap. We will consider the anisotropic trap later. Expectation of the effective Hamiltonian operator for our system (in the mean field level) is \cite{T-F}
\begin{eqnarray}
E=\sum_{\sigma=\uparrow,\downarrow}\int\int\epsilon_{\textbf{p},\textbf{r}}^{(0)}\bar{n}_\sigma(\textbf{p},\textbf{r})\frac{\text{d}^3\textbf{p}\text{d}^3\textbf{r}}{(2\pi\hbar)^3}+g\int\bar{n}_{\sigma}(\textbf{r})\bar{n}_{\sigma'}(\textbf{r})\text{d}^3\textbf{r},
\end{eqnarray}
where $\epsilon_{\textbf{p},\textbf{r}}^{(0)}=\frac{p^2}{2m}+\frac{1}{2}m\omega^2r^2$, $m$ is the mass of each particle, $\omega$ is the angular frequency of oscillations, $g=\frac{4\pi\hbar^2a_s}{m}$ is the coupling constant for the interparticle interaction ($g\delta^3(\textbf{r}-\textbf{r}')$  \cite{T-F,coupling}), $a_s$ is the s-wave scattering length, $\sigma'$ is the complementary of spin $\sigma$, $\bar{n}_\sigma(\textbf{r})=\int\bar{n}_\sigma(\textbf{p},\textbf{r})\text{d}^3\textbf{p}/(2\pi\hbar)^3$ is the averaged number density of spin $\sigma$ particles in equilibrium, and $\bar{n}_\sigma(\textbf{p},\textbf{r})$ is the average number of spin $\sigma$ particles in equilibrium at the single particle state $\{\textbf{p},\textbf{r}\}$. This number according to the Fermi-Dirac statistics is given by
\begin{eqnarray}
\bar{n}_\sigma({\textbf{p},\textbf{r}})=\frac{1}{\text{e}^{(\epsilon_{\textbf{p},\textbf{r}}-\mu_\sigma)/kT}+1},
\end{eqnarray}
where $\mu_\sigma$ is the chemical potential for the weakly interacting fermions, $\epsilon_{\textbf{p},\textbf{r}}=\frac{p^2}{2m}+V_{eff}(\textbf{r})$ is the mean field energy per particle, $V_{eff}(\textbf{r})=\frac{1}{2}m\omega^2r^2+g\bar{n}_{\sigma^{'}}(\textbf{r})$ is the effective mean field potential for a spin $\sigma$ particle \cite{giorgini}. It is to be mentioned that zero point energy can not alter the number distribution of particles as any constant shift in $\epsilon_{\textbf{p},\textbf{r}}$ can always be absorbed by $\mu_\sigma$. It is also to be mentioned that we can write Eqn.(2) only in the thermodynamic limit: $N_\sigma\rightarrow\infty$, $\omega\rightarrow0$ $\&$ $N_\sigma\omega^3=\text{constant}$, where $N_\sigma$ stands for total averaged number of particle having spin $\sigma$. In this limit, any integral over the discrete levels $\{n\}$ (of the oscillator energy: $(n+3/2)\hbar\omega$) is the same as that over the phase space ($\{\textbf{p},\textbf{r}\}$). Integrating $\bar{n}_\sigma({\textbf{p},\textbf{r}})$ in Eqn.(2) we get
\begin{eqnarray}
\bar{n}_\sigma(\textbf{r})=\frac{1}{\lambda_T^3}f_{3/2}(z_\sigma\text{e}^{-V_{eff}(\textbf{r})/kT}),
\end{eqnarray}
where $z_\sigma=\text{e}^{\mu_\sigma/kT}$ is the fugacity and $f_{j}(\text{x})=\text{x}-\frac{\text{x}^2}{2^{j}}+\frac{\text{x}^3}{3^{j}}-...$ is a Fermi function (or integral). In mathematics literature it is known as polylog function ($-Li_j(\text{-x})$) of order $j$. The chemical potential (or fugacity) is to be obtained from the constraint that
\begin{eqnarray}
N_\sigma=\int\bar{n}_{\sigma}({\textbf{r}})\text{d}^3\textbf{r}.
\end{eqnarray}
In an ideal situation ($g=0$), $\bar{n}_\sigma({\textbf{p},\textbf{r}})$ and $z_\sigma$ in Eqn.(2) become $\bar{n}_\sigma^{(0)}({\textbf{p},\textbf{r}})$ and $z_\sigma^{(0)}$ respectively. Thus for $g\rightarrow0$ Eqn.(4) becomes
\begin{eqnarray}
N_\sigma=\bigg(\frac{kT}{\hbar\omega}\bigg)^3f_3(z^{(0)}_{\sigma}).
\end{eqnarray}
\subsection{Shift in chemical potential due to the interaction}
Let the shift in chemical potential due to interaction be given by $\delta\mu_\sigma^{(\text{i})}=\mu_\sigma-\mu_\sigma^{(0)}$. This shift can now be obtained by the Taylor expansion of $\bar{n}_\sigma(\textbf{r})$ in Eqn.(3) about $g=0$ by considering $\delta z_\sigma^{(\text{i})} (=\text{e}^{\delta\mu^{(\text{i})}_\sigma/kT})$ an implicit function of $g$. Now, to the first order in $g$ and $\delta\mu^{(\text{i})}_\sigma$ we can write
\begin{eqnarray}
\bar{n}_\sigma(\textbf{r})=\bar{n}^{(0)}_\sigma(\textbf{r})&+&\frac{g\bar{n}^{(0)}_{\sigma}(\textbf{r})}{kT\lambda_T^3}f_{1/2}(z_\sigma\text{e}^{-\frac{m\omega^2r^2}{2kT}})\nonumber\\&+&\frac{\delta z^{(\text{i})}_\sigma}{z_\sigma\lambda_T^3}f_{1/2}(z_\sigma\text{e}^{-\frac{m\omega^2r^2}{2kT}}),
\end{eqnarray}
where $\bar{n}^{(0)}_\sigma(\textbf{r})$ is the number density for $g=0$. For a given $N_\sigma$, integrations of $\bar{n}_\sigma(\textbf{r})$ and $\bar{n}^{(0)}_\sigma(\textbf{r})$ in Eqn.(6) over $\textbf{r}$ are the same. Thus we get
\begin{eqnarray}
\int\bigg[\frac{\delta z_\sigma^{(\text{i})}}{z_\sigma\lambda_T^3}+\frac{g\bar{n}^{(0)}_{\sigma'}(\textbf{r})}{kT\lambda_T^3}\bigg]f_{1/2}(z_\sigma\text{e}^{-\frac{m\omega^2r^2}{2kT}})d^3\textbf{r}=0.
\end{eqnarray}
Now, to the first order in $g$ and $\delta z^{(\text{i})}_\sigma$ this shift is given by
\begin{eqnarray}
\frac{\delta z^{(\text{i})}_\sigma}{z_\sigma}=-\frac{g}{kT\lambda_T^3f_2(z_\sigma)}\sum_{i,j=1}^{\infty}\frac{(-1)^{i+j}z_{\sigma'}^{i}z_{\sigma}^{i}}{(i+j)^{3/2}i^{3/2}j^{1/2}}.
\end{eqnarray}
From Eqn.(8) we can also say that chemical potential (or Fermi energy as well) decreases due to the repulsive interparticle interaction among the particles. This is not a surprize as because increase of energy due to repulsive interaction does not necessarily mean chemical potential to be increased. We will give qualitative argument in this regard in section-VI. However, it should be mentioned that the Taylor expansion in Eqn.(6) is applicable only for weakly interacting regime ($a_s\bar{n}_{\sigma}^{1/3}(0)\ll1$) and not for strongly interacting regime ($a_s\bar{n}_{\sigma}^{1/3}(0)\gtrsim1$).

In this section we have highlighted the effect of weak interaction not only within mean field approximation but also within local density approximation which is embedded in effective mean field potential \cite{leblanc}. We will evaluate the interaction energy per particle later in section-IV. All the calculations in this section has also been done within semiclassical approximation due to the fact that we have integrated over phase space (from the beginning) instead of summing over discrete single particle levels \cite{salasnich}. This semiclassical approximation is valid only in the thermodynamic limit. As our system does not approach the thermodynamic limit, there must be an error of replacement of the sum by the integration. In the following we will incorporate this error as a finite size correction. We will also discuss other features of finite size correction as well.
\section{Finite size effect}
Apart from the replacement of the summation over the discrete levels ($\{n\}$) by the integration over ($\{n\}$) or that over the phase space ($\{\textbf{p},\textbf{r}\}$), finite size correction (to total energy) may also arise from the zero point energy of all the oscillators ($\sum_{\sigma=\uparrow,\downarrow}N_\sigma\frac{3}{2}\hbar\omega$) which should now be included in Eqn.(1) \cite{finitesize-fermi}. This replacement of summation, however, would impose a correction in density of states, such that a surface part (by name) $\frac{3}{2}\frac{\text{d}^2\textbf{p}\text{d}^2\textbf{r}}{(2\pi\hbar)^2}$ up to the first order in $\omega$ is to be added with the bulk part ($\frac{\text{d}^3\textbf{p}\text{d}^3\textbf{r}}{(2\pi\hbar)^3}$). The bulk and surface parts of course correspond respectively to the first and second terms of the degeneracy ($n^2/2+3n/2+1$) of the $n$th level of single particle energy $\epsilon_n=(n+\frac{3}{2})\hbar\omega$ \cite{biswas}. These are the two common origins of obtaining finite size correction. In the following we are going to introduce two more new origins.
\subsection{Corrections in temperature and chemical potential}
Macrostate of our system in thermodynamic equilibrium is usually described by the set $\{\omega,T,\mu_\sigma\}$, where $T$ and $\mu_\sigma$ (according to the zeroth law) are characteristics of an heat and particle reservoir. For our finite system, temperature is not measured directly \cite{temperature}. It is measured from the momentum distribution of atoms obtained after free expansion switching the magnetic trap off \cite{temperature}. The measured distribution is fitted with the momentum distribution formula $\bar{n}_\sigma(\textbf{p})=\frac{1}{(m\lambda_T\omega)^3}f_{3/2}(z_\sigma\text{e}^{-p^2/2mkT})$ which can be obtained by integrating $\bar{n}_\sigma(\textbf{p},\textbf{r}$) over $\textbf{r}$ \cite{craman}. This fitting formula is true only in the thermodynamic limit, and in the classical limit, its width is proportional to $\sqrt{T}$. Effect of zero point motion, however, has not been included in this fitting formula. Inclusion of this would certainly increase the width and the temperature as well. Chemical potential on the other hand, is not usually measured \cite{measurement}, but obtained from Eqn.(5) which is also true in the thermodynamic limit. Now we propose that actual temperature $T^{(\text{f})}$ and chemical potential ($\mu_\sigma^{(\text{f})}$) of a finite system (like ours) are different from the measured and calculated value respectively, and they must be the same in the thermodynamic limit.

Interparticle interaction may also change the momentum distribution, and we are considering its effect as a shift in fugacity and not as a shift in temperature. Thus apart from a finite size correction part ($\delta\mu_\sigma^{(\text{f})}$), $\mu_\sigma^{(\text{f})}$ also has an interaction correction part ($\delta\mu^{(\text{i})}_\sigma$) which has already been evaluated in Eqn.(8). We expect that $\delta T^{(\text{f})}=T^{(\text{f})}-T$ would be positive for smaller confinement (i.e. for tighter trap), and $\delta \mu_\sigma^{(\text{f})}=\mu_\sigma^{(\text{f})}-\mu_\sigma$ would be negative for a given $N_\sigma$. Now for $g=0$, Eqn.(4) along with the integration of $\bar{n}_\sigma(\textbf{p},\textbf{r})$ (in Eqn.(2)) over the surface part of the density of states ($\frac{3}{2}\frac{d^2\textbf{p}d^2\textbf{r}}{(2\pi\hbar)^2}$) becomes \cite{finitesize,biswas}
\begin{eqnarray}
N_\sigma=\bigg(\frac{kT^{(\text{f})}}{\hbar\omega}\bigg)^3f_3(z^{(\text{f,0})}_{\sigma})+\frac{3}{2}\bigg(\frac{kT^{(\text{f})}}{\hbar\omega}\bigg)^2f_2(z^{(\text{f,0})}_{\sigma})
\end{eqnarray}
where $z_\sigma^{(\text{f,0})}=z_\sigma^{(\text{0})}+\delta z_\sigma^{(\text{f})}$ is the actual fugacity for $g=0$. Let us further assume that $T^{(\text{f})}=T$ only at the absolute zero as for $T^{(\text{f})}\rightarrow0$, experimentalists may note that the width of momentum distribution is the minimum, and may set the measured temperature as $T=0$. Comparing Eqns.(5) and (9) we get the finite size correction in chemical potential for $T=0$, as $\delta\mu_\sigma^{(\text{f})}=-\frac{3}{2}\hbar\omega$. Henceforth we assume that $\delta\mu_\sigma^{(\text{f})}$ is the same in all temperature as commonly assumed for Bose gas particularly below the condensation point ($T_c$) \cite{finitesize}. This of course implies that $\delta z_\sigma^{(\text{f})}=z_\sigma^{(\text{f,0})}-z_\sigma^{(0)}=-z_\sigma^{(0)}3\hbar\omega/2kT$. Thus keeping $N_\sigma$ in Eqn.(9) fixed, we get the shift in temperature in terms of $\delta z_\sigma^{(\text{f})}$ as
\begin{eqnarray}
\frac{\delta T^{(\text{f})}}{T}=-\frac{\delta z_\sigma^{(\text{f})}}{3z^{(0)}_\sigma}\frac{f_2(z_\sigma)}{f_3(z_\sigma)}+{\it{O}}\bigg(\frac{\hbar\omega}{kT}\bigg)^2=\frac{1}{2}\frac{\hbar\omega}{kT}\frac{f_2(z_\sigma)}{f_3(z_\sigma)}.
\end{eqnarray}
For Bose gas, below $T_c$, $\delta z_\sigma^{(\text{f})}$ as well as $\delta T^{(\text{f})}$ has been assumed to be zero for obtaining a sharp condensation point and a considerable shift in $T_c$ \cite{finitesize,biswas}. But, in reality, a finite system of Bose gas does not have a sharp condensation point \cite{griesmaier}. Considering the same for Fermi gas, authors of Ref.\cite{finitesize-fermi} also did not get any considerable finite size correction in energy. That is why we have assumed $\delta z_\sigma^{(\text{f})}$ as well as $\delta T^{(\text{f})}$ to be nonzero. Our assumptions do not give a sharp condensation point of a Bose gas, and on the other hand, it may considerably improve the ideal gas prediction for our ultra-cold Fermi gas \cite{jin}.
\subsection{Correction in energy with anisotropy}
Let us now obtain the total energy of the system. In an ideal situation ($g=0$), Eqn.(1) along with the zero point energy and the integration over the surface part of the density of states gives the energy of spin $\sigma$ particles up to the first order in $\omega$ as
\begin{eqnarray}
E_\sigma^{(\text{f,0})}=\hbar\omega\bigg[3\bigg(\frac{kT^{(\text{f})}}{\hbar\omega}\bigg)^4f_4(z^{(\text{f,0})}_{\sigma})+\frac{9}{2}\bigg(\frac{kT^{(\text{f})}}{\hbar\omega}\bigg)^3f_3(z^{(\text{f,0})}_{\sigma})\bigg].
\end{eqnarray}
So far we have considered our system to be isotropic ($\omega_x=\omega_y=\omega_z=\omega$), and we already have noticed that tighter trap (larger $\omega$) would give a larger finite size correction. Finite size corrections for different values of $\omega_x, \omega_y, \omega_z$ would be larger than that for $\omega_x=\omega_y=\omega_z$ due to the fact that anisotropy results smaller (tighter) confinement and may even reduce the dimensionality eg. for $\omega_z\rightarrow\infty$ our 3-D trap may become a 2-D one. However, the prescription of including anisotropy (to the first order) is to replace $\omega$ in the finite size correction parts by its arithmetic mean $\bar{\omega}$, and in the bulk parts by its geometric mean ($\tilde{\omega}$) \cite{pitaevskii-rmp1}. Thus we recast Eqn.(11) along with Eqns.(10) and (5), and get energy per particle in terms of measured temperature as
\begin{eqnarray}
U_\sigma^{(\text{f})}=3kT\frac{f_4(z^{(0)}_{\sigma})}{f_3(z^{(0)}_{\sigma})}\bigg[1+\frac{\hbar\bar{\omega}}{kT}\bigg(2\frac{f_3(z^{(0)}_{\sigma})f_3(z^{(0)}_{\sigma})}{f_4(z^{(0)}_{\sigma})f_2(z^{(0)}_{\sigma})}-\frac{3}{2}\bigg)\bigg].
\end{eqnarray}

The inclusion of finite size effects along with the interaction effects have be done as separate corrections as all of their effects in our case are perturbative with respect to the unperturbed part (i.e. first term) of the expectation of Hamiltonian operator in Eqn.(1). Finite size effect can not, of course, be treated separately for strongly interacting regime.
\section{Evaluation of interaction energy}
Interaction energy per particle up to the first order in $g$ can be obtained from Eqn.(1) and (5). It is of course difficult to be evaluated for the presence of different $z^{(0)}_{\sigma}$ and $z^{(0)}_{\sigma'}$ in $\bar{n}^{(0)}_{\sigma}(\textbf{r})$ and $\bar{n}^{(0)}_{\sigma'}(\textbf{r})$ respectively. At this stage we need to evaluate $\frac{\bar{n}^{(0)}_{\sigma'}(\textbf{r})}{\bar{n}^{(0)}_{\sigma}(\textbf{r})}$ specially for weak interaction and small number fluctuation ($\frac{N_{\sigma}-N_{\sigma'}}{N_{\sigma}}=\frac{\triangle N_{\sigma}}{N_{\sigma}}\ll1$). While for $T/T_F\rightarrow\infty$, we obviously have the classical result: $\frac{\bar{n}^{(0)}_{\sigma'}(\textbf{r})}{\bar{n}^{(0)}_{\sigma}(\textbf{r})}\rightarrow\frac{N_{\sigma'}}{N_{\sigma}}$; for $T/T_F\rightarrow0$, we have $\frac{\bar{n}^{(0)}_{\sigma'}(\textbf{r})}{\bar{n}^{(0)}_{\sigma}(\textbf{r})}\rightarrow\sqrt{\frac{N_{\sigma'}}{N_{\sigma}}}\big(1-{\it{O}\big(\frac{m\omega^2r^2}{4kT_F}\frac{\triangle N_\sigma}{N_\sigma}\big)}\big)$ within Thomas-Fermi approximation \cite{pitaevskii-rmp2}. Although for small number fluctuation, these two limits do not differ ($\sim\frac{\triangle N_\sigma}{2N_\sigma}$) appreciably, yet we stick with the Thomas-Fermi result as because weak interaction plays role only for $T/T_F\ll1$. Thus for weak interaction and small number fluctuation, we give an \textit{ansatz}: $\frac{\bar{n}^{(0)}_{\sigma'}(\textbf{r})}{\bar{n}^{(0)}_{\sigma}(\textbf{r})}\approx\sqrt{\frac{N_{\sigma'}}{N_{\sigma}}}$ so that the interaction energy can be evaluated in a closed form. This \textit{ansatz}, of course, is not applicable for strongly interacting regime and large number fluctuation. Now, with this \textit{ansatz}, we get interaction energy per spin $\sigma$ particle as
\begin{eqnarray}
U^{(\text{int})}_\sigma=\frac{g}{\lambda_{T}^3}\sqrt{\frac{N_{\sigma'}}{N_\sigma}}\bigg(\frac{h(z^{(0)}_{\sigma})}{f_3(z^{(0)}_{\sigma})}\bigg),
\end{eqnarray}
where
\begin{eqnarray} h(z_\sigma^{(0)})=\frac{4}{\sqrt{\pi}}\int_0^\infty\big[f_{\frac{3}{2}}(z_\sigma^{(0)}\text{e}^{-\text{x}^2})\big]^2\text{x}^2\text{d}\text{x}.
\end{eqnarray}
One check that for $T\gtrsim T_F$, the integral representation of $h(z_\sigma^{(0)})$ in Eqn.(14) can be approximated by the closed form $\frac{1}{2^{3/2}}\big[f_{9/4}(z_\sigma^{(0)})\big]^2$.
\section{Corrected excess energy}
Energy per particle ($U_\sigma$) to the first order in $\bar{\omega}$ and $g$ can be obtained as a summation of its two parts in Eqn.(12) and (13). It is easy to check that its classical part is $U_\sigma^{\text{cl}}=3kT$. Thus we get excess energy per particle ($\delta U_\sigma=U_\sigma-U_{\sigma}^{(\text{cl})}$) for our anisotropic system as
\begin{eqnarray}
\frac{\delta U_\sigma}{U_{\sigma}^{(\text{cl})}}&=&\frac{f_4(z^{(0)}_{\sigma})}{f_3(z^{(0)}_{\sigma})}\bigg[1+\frac{\bar{\omega}}{\tilde{\omega}}\bigg(\frac{T_F/T}{(6N_\sigma)^{1/3}}\bigg)\bigg\{2\frac{f_3(z^{(0)}_{\sigma})}{f_4(z^{(0)}_{\sigma})}\frac{f_3(z^{(0)}_{\sigma})}{f_2(z^{(0)}_{\sigma})}\nonumber\\&&-\frac{3}{2}\bigg\}+\frac{g}{3kT\lambda_{T}^3}\sqrt{\frac{N_{\sigma'}}{N_\sigma}}\frac{h(z^{(0)}_{\sigma})}{f_4(z^{(0)}_{\sigma})}\bigg]-1,
\end{eqnarray}
where $kT_F$ gets the value $\hbar\tilde{\omega}(6N_\sigma)^{1/3}$ \cite{pitaevskii-rmp2}.
\subsection{Chemical potential in a closed form of temperature}
If we want to plot the excess energy (in Eqn.(15)) with respect to $T$, we need to know the temperature dependence of $\mu_\sigma^{(0)}$ from Eqn.(5). But, it is impossible to be known as the inverse of a polylog function does not exist. There is no other way but to get $\mu_\sigma^{(0)}$ in a closed form without an approximation. However, whether an approximate $\mu_\sigma^{(0)}$ is good can be justified from the exact graphical solutions which might have been used by DeMarco and Jin for plotting the ideal gas prediction \cite{jin}. In the following we prescribe an approximate method for obtaining a good $\mu_\sigma^{(0)}$ in a closed form relevant for the entire range of temperature.

Let us first investigate the approximate $\mu_\sigma^{(0)}(T)$ in the degenerate regime ($T/T_F\ll1$). For $\frac{T}{T_F}\rightarrow0$, Fermi functions follow Sommerfeld's asymptotic formula \cite{pathria}
\begin{eqnarray}
f_j(z_\sigma^{(0)})=\frac{[\mu^{(0)}_\sigma/kT]^j}{\Gamma(j+1)}\bigg[1+\frac{\pi^2}{6}\frac{j(j-1)}{[\mu^{(0)}_\sigma/kT]^2}+...\bigg].
\end{eqnarray}
Now for $j=3$, Eqns.(5) and (16) lead to the asymptotic form of chemical potential
\begin{eqnarray}
\mu_\sigma^{(0)}(T_<)=kT_F\bigg[1-\frac{\pi^2}{3}\bigg(\frac{T}{T_F}\bigg)^2+...\bigg].
\end{eqnarray}
One can check that the approximate chemical potential in Eqn.(17) is reasonably good for $T/T_F\lesssim0.5$.

\begin{figure}
\includegraphics{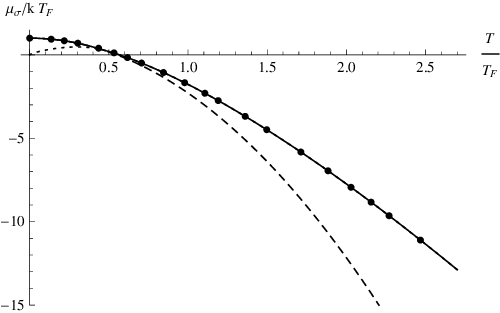}
\caption {Dashed and dotted lines represent Sommerfeld's result (Eqn. (17)) and improved classical result (Eqn.(19)) respectively. Solid line represents the approximate chemical potential in Eqn.(20). Points represent exact graphical solutions of $\mu_\sigma(T)$ in Eqn.(5).}
\end{figure}

Let us also investigate the approximate $\mu_\sigma^{(0)}(T)$ in the classical regime ($T/T_F\gg1$). Taking only the first term of the Taylor series expansion of $f_3(z_{\sigma}^{(0)})$ in Eqn.(5) we get the approximate chemical potential as
\begin{eqnarray}
\mu_\sigma^{(0)}(T_{\gg})=-3kT_F\frac{T}{T_F}\text{ln}\bigg(6^{1/3}\frac{T}{T_F}\bigg).
\end{eqnarray}
Although the approximate $\mu_{\sigma}^{(0)}$ in Eqn.(18) is valid for $T/T_F\rightarrow\infty$ yet one can check that it is reasonably good even for $T/T_F\gtrsim1.5$.

However, in DeMarco and Jin's experiment \cite{jin}, observation of quantum degeneracy was highlighted for $0.5\lesssim T/T_F\lesssim1.5$. In this regime none of the above approximations is good. In the next, we are going present a calculation of obtaining a more accurate form of the chemical potential relevant not only for $0.5\lesssim T/T_F\lesssim1.5$ but for the entire range of temperature.

Classical result in Eqn.(18) can be improved by considering the first three terms of the Taylor series expansion of $f_3(z_{\sigma}^{(0)})$. Real root for the cubic Eqn. of $z_\sigma^{(0)}$ now leads to
\begin{eqnarray}
\mu_\sigma^{(0)}(T_{>})=kT_F\frac{T}{T_F}\text{ln}\bigg[\frac{9}{8}-\frac{165\times3^{1/3}f(t)}{8}+\frac{3^{2/3}}{8f(t)}\bigg],
\end{eqnarray}
where $f(t)=t/[128-783t^3+16\sqrt{64-783t^3+8244t^6}]^{1/3}$ and $t=T/T_F$. Sommerfeld's result (Eqn.(17)) and the improved classical result (Eqn.(19)) intersects at $t\approx0.468$. Thus approximate chemical potential for $0\le\frac{T}{T_F}<\infty$ can be written from the combination of these two as
\begin{eqnarray}
\mu_\sigma^{(0)}(T)=\mu_\sigma^{(0)}(T_{<})\theta(0.468-t)+\mu_\sigma^{(0)}(T_{>})\theta(t-0.468).~~
\end{eqnarray}
Comparing with the exact graphical solutions in FIG. 1 we can say that approximate $\mu_{\sigma}^{(0)}$ in Eqn.(20) is reasonably good for the entire range of temperature.

\begin{figure}
\includegraphics{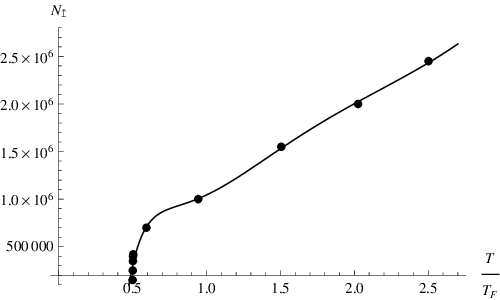}
\caption {A few experimental data points having $50\%$ uncertainty are adapted from Ref.\cite{jin}. Solid line represents a possible fitting (Eqn.(21)) for the data points.}
\end{figure}

\subsection{Data fitting for evaporation of particles}
However, for comparing our analytic result with the experimental data, we need to know the temperature dependence of the number of particles that arises due to evaporation of particles during the experiment of DeMarco and Jin \cite{jin}. Although the average number of particles is not a constant yet they compared their experimental data with the ideal gas prediction. Ideal gas prediction for a fixed $N_\sigma$ remains unaltered if $N_\sigma$ changes quasi-statistically with temperature. Thus temperature dependence of $N_\sigma$ in our theory would come only in the corrections parts in Eqn.(15).

DeMarco and Jin showed the temperature dependence of $N_\uparrow$ in a figure keeping $N_\uparrow:N_\downarrow=60\%:40\%$ unchanged within $5\%$ error during the evaporation ramp \cite{jin}. We replot their data in FIG. 2, and a possible fitting for these data points can be obtained with the help of Wolfram Mathematica, and it is given by
\begin{eqnarray}
N_{\uparrow}(t)=10^{11}(c_0+c_1t^{\frac{1}{16}}+c_2t^{\frac{1}{8}}+c_3t^{\frac{1}{4}}+c_4t^{\frac{1}{2}}+c_5t),~~
\end{eqnarray}
where $c_0=-2.05293$, $c_1=6.55557$, $c_2=-6.03520$, $c_3=1.69586$, $c_4=-0.168018$, $c_5=0.00472971$. From FIG. 2 we can see that $N_\uparrow$ varies from $\sim2.7\times10^6$ to $\sim4.2\times10^5$ as $T/T_F$ decreases from 2.7 to 0.5. Multiplying the right hand side of Eqn.(21) by $2/3$ we get $N_{\downarrow}(t)$ as well.

Apart from the explicit temperature dependence of the different parts in Eqn.(15), finite size correction part in this equation also has an implicit temperature dependence as DeMarco and Jin smoothly varied the radial trap frequency $\omega_\perp$ ($=\omega_x=\omega_y$) from $2\pi\times44$ Hz to $2\pi\times370$ Hz keeping the axial trap frequency fixed ($\omega_z=2\pi\times19.5$ Hz) for obtaining the temperature dependence of $N_\sigma$ \cite{jin}. Thus $\omega_\perp$ in $\bar{\omega}/{\tilde{\omega}}$ of Eqn.(15) has an implicit temperature dependence which was not shown in Ref.\cite{jin}. However, for a quasi-statically smooth variation of $\omega_\perp$, one can expect the ratio of arithmetic and geometric means of $\omega_x, \omega_y$ and $\omega_z$ to be a constant around $T/T_F=1$, and we set $\bar{\omega}/\tilde{\omega}=1.3677$ by picking up $\omega_\perp=2\pi\times137$ Hz for $T/T_F\approx1$ \cite{jin}.

\begin{figure}
\includegraphics{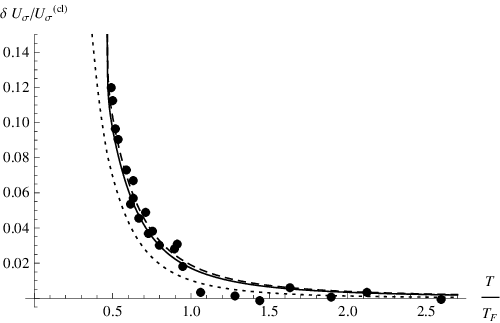}
\caption {Dotted line represents ideal gas prediction \cite{jin}. Solid and dashed lines represent our result in Eqn.(15) for spin $\uparrow$ and $\downarrow$ respectively. Several experimental data points having $20\%$ uncertainty are adapted from Ref.\cite{jin}. Here we take $m=40\times1.675\times10^{-27}$ \text{kg}, $\bar{\omega}/\tilde{\omega}=1.3677$, $N_\uparrow:N_\downarrow=3:2$ \cite{jin}, and $a_s=169 a_0$ \cite{jin1}.}
\end{figure}

\subsection{Our result for excess energy per particle}
We replace $N_\sigma$ in different parts of Eqn.(15) by $N_\sigma(t)$ in Eqn.(21), and plot the excess energy in Eqn.(15) (along with the approximate bulk chemical potential in Eqn.(20)) in FIG. 3 for spin $\uparrow$ and spin $\downarrow$ particles keeping $\bar{\omega}/\tilde{\omega}$ unchanged with its value at $T\approx T_F$. In the same figure we also plot the ideal gas prediction putting $N_\sigma=\infty$ and $g=0$ in Eqn.(15). From this figure we can say that our analytic result matches better with the experimental data.
\section{Comparisons among ideal and correction parts}
As a consistency check and for showing plausibility of our theory, let us now compare the interaction energy and potential energy for different temperatures. It is easy to understand from Eqn.(13) that $N_{\sigma}U_{\sigma}^{(\text{int})}$ is the total interaction energy ($E^{(\text{int})}$). We can also say from the virial theorem that $(N_{\sigma}U_{\sigma}^{(0)}+N_{\sigma'}U_{\sigma'}^{(0)})/2$ is the total oscillator energy ($E^{(\text{ho})}$). Thus from Eqn.(15), we get their ratio for $N_\sigma=N_{\sigma'}=N/2$ as
\begin{eqnarray}
\frac{E^{(\text{int})}}{E^{(\text{ho})}}=\bigg(\frac{g}{3kT\lambda_T^3}\bigg)\frac{h(z_\sigma^{(0)})}{f_4(z_\sigma^{(0)})}+{\it{O}}(g^2).
\end{eqnarray}
We plot the right hand side of Eqn.(22) for a fixed number of particles in FIG. 4. However, to obtain  $\frac{E^{(\text{int})}}{E^{(\text{ho})}}$ for $T\rightarrow0$, we must need to know the asymptotic expansions of $f_4(z_\sigma^{(0)})$ and $h(z_\sigma^{(0)})$. While the first one is followed from Eqn.(16) second one is to be obtained from Eqn.(14). In the integral representation of $h(z_\sigma^{(0)})$, the Fermi function contributes essentially for $\text{x}<(\text{ln} z_\sigma^{(0)})^{1/2}$. Now from Eqns.(14) and (16) we get the desired asymptotic expansion as
\begin{eqnarray}
h(z_\sigma^{(0)})=\frac{1024\big(\text{ln} z_\sigma^{(0)}\big)^{\frac{9}{2}}}{2835\pi^{3/2}}\bigg[1+\frac{21\pi^2}{32\big(\text{ln} z_\sigma^{(0)}\big)^{2}}+...\bigg].
\end{eqnarray}
This form has been used in FIG. 4 in particular for $T/T_F\le0.1$. Thus for $T\rightarrow0$, Eqn.(22) gives
\begin{eqnarray}
\frac{E^{(\text{int})}(0)}{E^{(\text{ho})}(0)}&=&\frac{8192\times\sqrt{2}\times3^{1/6}}{2835\pi^2}\bigg(\frac{N^{1/6}a_s}{\tilde{l}}\bigg)\nonumber\\&\approx&0.497\bigg(\frac{N^{1/6}a_s}{\tilde{l}}\bigg),
\end{eqnarray}
where $\tilde{l}=\sqrt{\frac{\hbar}{m\tilde{\omega}}}$ is the confining length scale of a particle in our anisotropic trap. One can check that Eqn.(24) is consisted with the Thomas-Fermi result for $T\rightarrow0$ \cite{T-F}.

\begin{figure}
\includegraphics{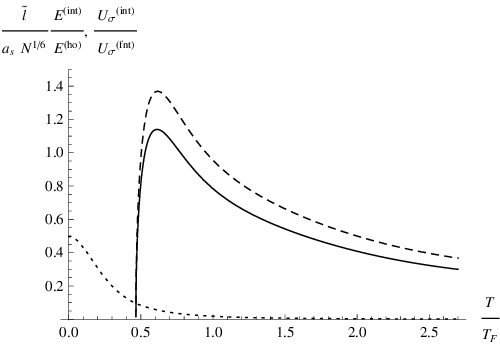}
\caption {Dotted line represents the ratio of interaction and potential energy (Eqn.(22)) for fixed number of particles in units of $a_sN^{1/6}/\tilde{l}$. Solid and dashed lines represent the ratio of interaction and finite size corrections (in Eqn.(15)) for spin $\uparrow$ and $\downarrow$ particles respectively.}
\end{figure}

As a corollary of the asymptotic expansion, we can also calculate the shift in Fermi energy (from Eqn.(8)) due to interaction as
\begin{eqnarray}
\frac{\delta\mu^{(\text{i})}_\sigma(0)}{\mu_\sigma(0)}&=&-\frac{128\times\sqrt{2}\times3^{1/6}}{105\pi^2}\bigg(\frac{N^{1/6}a_s}{\tilde{l}}\bigg)\nonumber\\&\approx&-0.210\bigg(\frac{N^{1/6}a_s}{\tilde{l}}\bigg).
\end{eqnarray}
That Fermi energy (or Fermi temperature as well) decreases in Eqn.(25) is not a surprise as because the number density of fermions at the central region of the trap decreases for repulsive interaction. It is a similar phenomenon of how the condensation temperature of a harmonically trapped Bose gas decreases for repulsive interaction \cite{pitaevskii-rmp1}. It should also be mentioned in this regard that Pauli paramagnetism arises for $\mu_\downarrow(0)>\mu_\uparrow(0)$ while $U_\uparrow(0)>U_\downarrow(0)$ \cite{pathria}.

It is not clear from FIG. 3 how big is the finite size correction with respect to the interaction correction. Dividing $U_\sigma^{(\text{int})}$ by $U_\sigma^{(\text{fnt})}$ in Eqn.(15) we get the ratio of the interaction and finite size correction, and plot it in FIG. 4 for spin $\uparrow$ and $\downarrow$ particles with the experimental parameters used in FIG 3. Now we can say, from FIG. 4, that interaction and finite size corrections are comparable within the temperature range of our interest. This comparison for $T/T_F<0.5$, however, is ill defined as the temperature dependence of $N_\sigma$ is insensitive for $T/T_F\lesssim0.5$ in FIG. 2.
\section{Conclusion}
We have presented an analytic calculation for exploring thermodynamic behavior of a harmonically trapped weakly interacting Fermi gas through the temperature dependent energy. Our result for $\frac{T}{T_F}\rightarrow0$ and $\frac{T}{T_F}\gg1$, reproduces the standard Thomas-Fermi \cite{T-F} and classical results \cite{jin} respectively, and of course matches better with the experimental data \cite{jin}.

Since the number density of the fermions in the trap is reasonably low ($\sim10^{16}/\text{m}^3$) one may expect ladder approximation as a better choice with respect to our choice for the mean field (Hartree-Fock) approximation \cite{fetter}. However, we can not get any contribution from the ladder diagrams if we consider short ranged interaction ($g\delta^3(\textbf{r}-\textbf{r}')$) among the particles. Hence, Hartree-Fock approximation is reasonably good even for the low dense gas considered by us.

Finite size correction scheme prescribed here can also be generalized for the Bose gas. For Bose gas, Fermi functions ($f_j(\text{x})$) are to be replaced by the Bose-Einstein functions ($g_j(\text{x})$).

Approximate chemical potential (in a closed form) obtained by us makes an analytic theory plausible in the entire range of temperature.

Our main new result is the finite size correction part (which is first order in $(\frac{\hbar\bar{\omega}}{kT})\sim0.01$) in Eqn.(15). It has come from the considerations of (i) zero point energy, (ii) discreteness of single particle energy spectrum, (iii) shift in temperature, and (iv) shift in chemical potential. While first two considerations result correction to the second order in $(\frac{\hbar\omega}{kT})$ \cite{finitesize-fermi}, the last two considerations result to the first order in $\frac{\hbar\bar{\omega}}{kT}$. This first order correction along with interparticle interaction in Eqn.(15) may considerably improve the ideal gas prediction. Our result can, of course, be negligibly improved to the second order by including the result of Ref.\cite{finitesize-fermi}. Instead of taking an intermediate value of $\bar{\omega}/{\tilde{\omega}}$, our result can also be improved a little if one gets the data for the implicit temperature dependence of $\omega_\perp$ (or of $\bar{\omega}/{\tilde{\omega}}$) in Eqn.(15).

Excess energy in other experiments also found to be a little larger with respect to the ideal gas prediction \cite{jin1}. Our result (for the finite size and interaction corrections) are somewhat uncertain at the present level of accuracy \cite{jin, jin1}, and compelling the experimental evidences for its presence. Future experiment with reduced error may confirm our prediction.
\section*{Acknowledgment}
This work has been sponsored by the University Grants Commission [UGC] under the D.S. Kothari Postdoctoral Fellowship Scheme {[No.F.4-2/2006(BSR)/13-280/2008(BSR)]}. Useful comments of J.K. Bhattacharjee of SNBNCBS and K. Sengupta of IACS are gratefully acknowledged. Hospitalities of IACS and IISER-Kolkata are also acknowledged.

\end{document}